\documentclass{article}
\usepackage{graphicx}
\usepackage{amssymb}
\usepackage{amsmath}
\usepackage{tikz}
\usepackage{url}
\usetikzlibrary{shapes}

\title{Genome Compression Against a Reference}

\author{Aniruddha Laud
\and
Gaurav Menghani
\and
Madhava Keralapura\\ \\
Stony Brook University
}

\date{December 7, 2011}

\begin{document}

\maketitle
.

\vspace*{0.25cm}
\begin{abstract}
Being able to store and transmit human genome sequences is an important part in
genomic research and industrial applications. The complete human genome has 3.1
billion base pairs (haploid), and storing the entire genome naively takes about
3 GB, which is infeasible for large scale usage.

However, human genomes are highly redundant. Any given individual's genome would
differ from another individual's genome by less than 1\%. There are tools like
DNAZip, which express a given genome sequence by only noting down the
differences between the given sequence and a reference genome sequence. This
allows losslessly compressing the given genome to $\approx$ 4 MB in size.
 
In this work, we demonstrate additional improvements on top of the DNAZip
library, where we show an additional $\approx$ 11\% compression on top of
DNAZip's already impressive results. This would allow further savings in disk
space and network costs for transmitting human genome sequences.
\end{abstract}
\vspace*{0.25cm}

\clearpage

\tableofcontents

\clearpage

\section {Introduction}
Genomes are usually distributed in two formats. Firstly, using the FASTA format.
Secondly, and more commonly as two files, one containing the SNPs (Single
Nucleotide Polymorphisms), and the other containing insertions and deletions,
both with respect to a reference genome (usually hg18 
\cite{ucschg, ucschg18}).\\ 
\\
While the size of a pure FASTA-based genome is about 3 GB, using the second
format as mentioned above, James Watson's genome was expressed in about 1.8 GB
\cite{jwseq}, with reference to the hg18 genome. \\
\\
The DNAZip project \cite{dnazip} makes use of the database of all SNPs reported
(dbSNP) and the hg18 genome to represent this information cleverly in about 4.1
MB. We intend to represent this information in an even more efficient format,
and reduce the size of the genome further, without using any extra information.
\clearpage

\section {Relevant Work}
The following techniques are used by DNAzip to achieve compression.
\begin{enumerate}

\item Variable size integer representation and delta positions:\\
In the variation data, all the variation (indels and SNPs) are expressed as
position on the reference genome, plus the variation actually present. The
position is expressed as an offset from the beginning of the sequence. Thus, the
size required to store the position increases while storing variations at the
end of the genome. Instead of storing the entire offset, it is sufficient to
store just the delta from the previous variation. One problem with this is that
an integer value takes the same amount of space to store, regardless of the
value within it. 

Thus, by implementing a mechanism to store exactly as many bits are required,
storage utilized could be optimized. A variable integer as implemented by DNAzip
uses 7 of 8 bits in a character to store the bits of the integer, and 1 bit to
indicate whether an integer ends at that byte or not.\\

\item{SNP mapping}\\
Most of the variation between genomes are generally substitutions usually
referred to as single nucleotide polymorphisms (SNPs). Most of these
substitutions exist in one of two possible alternatives (bi-allele). Most of
these SNPs are collected and organized by NCBI in what is known as dbSNP. dbSNP
contains all known SNPs organized by position against the human reference genome
(hg18, for instance). Though dbSNP records variations other than just the SNPs
(it also records indels and multi-allele SNPs), DNAzip just considers the
bi-allele SNPs. DNAzip stores the SNPs as a bit vector of all possible SNPs,
with 1 where the SNP exists and 0 where the SNP doesn't. For those SNPs that are
not present in dbSNP, they are stored as position along with the actual
substitution.\\

\item{K-mer partitioning}\\
A common compression technique used is Huffman coding. By partitioning all the
insertion data into K-mers and inserting them into a Huffman tree, the optimal
representation of each k-mer can be obtain, and this representation is then
stored, thus achieving some level of compression.\\
\end{enumerate}
\clearpage

\section {Work Done}

\subsection {Storing the Bit Vectors Efficiently}
As explained above, DNAZip creates a bitmap from the dbSNP. If bit i is set in
the bitmap, it implies that the ith SNP from the dbSNP occurs in the actual
genome. The DNAZip project simply stores this sparse bitmap for each chromosome.
Upon analysis, we found that these bitmaps compose of about 1.2 MB of the total
4.1 MB of the genome. This sparseness of the bitmap made a strong case for
compressing it.

Compressing the bit-vectors using Huffman Encoding with k-mers of size 5
resulted in the genome being of size 3896467 bytes. While using k-mers of size 6
resulted in the genome being of size 3889100 bytes. This is against the original
file of size 4198717 bytes. Thus, using k-mers of size 6, we decreased the size
of the genome by 302.36 KB, which is a space-saving of 7.37\%.\\

\begin{table}
\begin{center}
	\begin{tabular}{|p{1in}|p{1in}|}

	\hline
	Chromosome	&		\% Compression Achieved \\
	\hline
1 & 26.4089\\
	\hline
2 & 21.779\\
	\hline
3 & 20.9808\\
	\hline
4 & 19.7785\\
	\hline
5 & 24.719\\
	\hline
6 & 25.1927\\
	\hline
7 & 24.6257\\
	\hline
8 & 22.4158\\
	\hline
9 & 32.0532\\
	\hline
10 & 23.3288\\
	\hline
11 & 22.3557\\
	\hline
12 & 22.7921\\
	\hline
13 & 17.8374\\
	\hline
14 & 22.389\\
	\hline
15 & 22.9731\\
	\hline
16 & 28.062\\
	\hline
17 & 26.9674\\
	\hline
18 & 22.1033\\
	\hline
19 & 27.468\\
	\hline
20 & 31.6912\\
	\hline
21 & 19.8773\\
	\hline
22 & 33.8321\\
	\hline
M & 35.4922\\
	\hline
X & 49.1172\\
	\hline
Y & 80.8371\\

	\hline
	\end{tabular}
\end{center}
	\caption{Compression achieved for each chromosome}

\end{table}

\clearpage

\subsection {Improving the Variable Integer (VINT) storage}
The DNAZip source code makes heavy use of Variable Integer (VINT) storage, so
that they do not have to allocate a fixed number of bits for each integer value
that needs to be written. The reason being, if we are writing very small values,
and the data type of our choice is the standard integer, we would be using 32
bits for each value that we write but most of the higher bits in the number
would not be set. However, if we choose a small data type like a byte, we would
not be able to write values greater than $2^8 - 1$. \\
This is solved by writing 8 bits at a time, of which the MSB is set if there is
another block of 8 bits for the number after the current block. Thus, out of the
8 bits, 1 bit is a flag, and the rest 7 come from the number to be written. We
realized that this might not be the most efficient way of doing it. A large
number of values written as variable integers are small delta values, and
writing 8 bits at a time would mean that atleast 8 bits would be written even if
the number is very small. \\
The values being written as VINTs during the compression process were written to
an auxilliary file. We then tried to compare the space consumed when the size of
the VINT word size is varied. 

\begin{figure}[htp]
\centering
\includegraphics[scale=0.4]{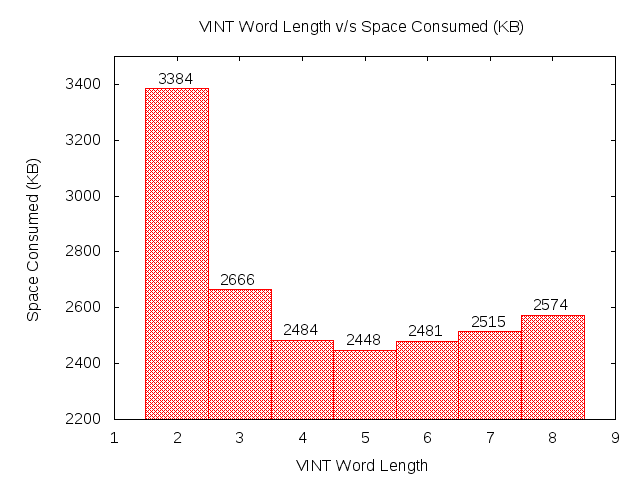}
\caption{VINT Word Length v/s Space Consumed (KB)}\label{fig:fs}
\end{figure}
\clearpage
Thus, with the VINT word size as 8, 2574 KB of space was being used up for
writing VINTs. The optimal VINT word size was found to be 5, which as per
experiments would have used up 2448 KB of space.\\
\\ 
The actual compression received from this was 129 KB roughly, which was in line
with our expectation before this experiment.\\
\clearpage

\subsection {Improving the storage of Insertions/Deletions (INDELs)}
After thorough analysis, we realized that the Insertion/Deletion format in
iteself was pretty packed. One optimization possible was to create a Bit Vector
for the Insertions and Deletions which were present in dbSNP. This meant that we
would not need to store the information for the insertions and deletions which
existed in the dbSNP. A bit vector for the insertions / deletions present in
dbSNP could be used to encode that information.\\ 
\\
The number of deletions in the genome which also existed in dbSNP, was
significant but not large enough (Refer Table 2). The percentage of deletions in
a chromosome which exist in dbSNP varies from 5.3\% to 11.68\% (apart from the
Mitochondrial DNA, which has only one deletion, which does not exist in
dbSNP).\\
\\
We compressed the bit vector using Run-Length Encoding. This gave us a saving of
about 10 KB. Using Huffman Encoding gave us a saving of roughly 20 KB. Another 1
KB of compression was squeezed in by noticing that the bit vector was sparser
than the SNP bit vector, and a k-mer length of size 7 was the most suitable for
this.\\ 
\\
Thus, we only changed the storage for the deletions. The deletions which were
present in dbSNP were stored using bitvectors, and compressed using Huffman
Encoding with k-mer size as 7.\\
\\
For insertions we noticed that the number of insertions in the Genome which
existed in dbSNP were very few. Thus, the bit vector was extremely sparse. And
hence, there was not much benefit of encoding them as bit-vectors.
\begin{table}[h]
\begin{center}

  	\begin{tabular}{|p{1in}|p{1in}|}
	
	\hline
	Chromosome	&		\% of Deletions found in dbSNP \\
	\hline
	1	&	8.54 \\
	\hline
	2	&	8.76 \\
	\hline
	3	&	8.80 \\
	\hline
	4	&	11.67 \\
	\hline
	5	&	10.46 \\
	\hline
	6	&	8.87 \\
	\hline
	7	&	6.90 \\
	\hline
	8	&	9.45 \\
	\hline
	9	&	7.84 \\
	\hline
	10	&	9.40 \\
	\hline
	11	&	10.09 \\
	\hline
	12	&	6.39 \\
	\hline
	13	&	6.70 \\
	\hline
	14	&	9.63 \\
	\hline
	15	&	6.99 \\
	\hline
	16	&	6.74 \\	
	\hline
	17	&	6.50 \\
	\hline	
	18	&	10.61 \\
	\hline	
	19	&	5.30 \\
	\hline
	20	&	13.06 \\	
	\hline
	21	&	10.36 \\
	\hline
	22	&	7.08  \\
	\hline
	Mitochondrial	&	0.0 \\
	\hline
	X	&	7.57 \\
	\hline
	\end{tabular}
\end{center}
  \caption {Percentage of Deletions in each chromosome found in dbSNP}
  \end{table}

\clearpage

\subsection {Miscellaneous Experiments}

\begin{enumerate}

\item For SNPs which are not present in dbSNP, DNAZip writes the substituted and
the substituted characters as bits to the file. We tried compressing this string
with different k-mer sizes but the bit-string was random enough not to yield any
decrease in the compression.

\item Further in our efforts to improve the encoding of SNPs which were not
present in dbSNP, we tried storing all the substitute characters in order of the
character they replace. However, now this increased the delta values, this did
not give a better compression. Efforts to tweak the VINT storage did not help
either.

\end{enumerate}

\clearpage

\section {Results \& Conclusion}
The original DNAZip format was already pretty efficient. However, over the
course of project, we have exploited avenues where we could foresee significant
compression possible. Towards the end of the project, it became hard to find any
further improvements.\\
\\
In all we have compressed James Watson's genome to \textbf{3736918} bytes, from
the original 4198717 bytes used by the DNAZip format. This is a compression of
roughly \textbf{11\%}. \\
\\
With increase in the size of dbSNP, we expect the compression to become better
with time, since majority of the file size, roughly 2 MB out of the current 3.56
MB is in deltas used for encoding SNPs absent from dbSNP. With time this number
is expected to go down, and the bit-vector size to increase. Also, the
bit-vector gets sparser and the compression is expected to work better.\\
\\
\clearpage

\section {Further Work}

\begin{enumerate}
\item After experimentation we have made certain choices, such as the k-mer size
for compressing the SNP bitvectors. While we expect other genomes to have
similar data and our choices to be relevant, but we can optimize this at the
cost of performance. Doing one pass to figure out the data to be written, such
as the bit-vectors, VINT data, and then trying out all possible choices and
making the best possible choice. However, as apparent, it would be roughly more
than 2x slower than the current DNAzip build with our code.

\item Of the 2574 KB occupied by VINT storage, roughly 2 MB is used by Deltas
stored while encoding SNPs. Using a different VINT storage as discussed earlier
we saved about 129 KB. However, we think there should be scope of further
optimizing the VINT storage. 

\end{enumerate}

It is possible to make the DNAZip utility more useful. A few of them are as
follows:

\begin{enumerate}
\item Allowing compression and decompression on a chromosome-by-chromosome
basis.

\item There is scope for constant-factor improvement in DNAzip's performance,
this could be done by using C-style data structures instead of the C++ STL data
structures being used.
\end{enumerate}

\end{document}